\documentclass[manuscript]{emulateapj}

\def\gsim{\mathrel{\raise0.35ex\hbox{$\scriptstyle >$}\kern-0.6em 
\lower0.40ex\hbox{{$\scriptstyle \sim$}}}}
\def\lsim{\mathrel{\raise0.35ex\hbox{$\scriptstyle <$}\kern-0.6em 
\lower0.40ex\hbox{{$\scriptstyle \sim$}}}}

\def\oii{{\rm [O{\sc ii}]}}

\def\oiii{{\rm [O{\sc iii}]}}
\def\nii{{\rm [N{\sc ii}]}}
\def\hii{{\rm H{\sc ii}}}

\def\leftangle{<}
\def\rightangle{>}

\slugcomment{ Accepted for publication in ApJ }

\shorttitle{Red Star Forming Galaxies and Their Environment at $z=0.4$}
\shortauthors{Y. Koyama et al.}

\begin{document}


\title{Red Star Forming Galaxies and Their Environment at $z=0.4$ \\
Revealed by Panoramic H$\alpha$ Imaging}


\author{Yusei Koyama$^{1}$, Tadayuki Kodama$^{2,3}$, 
Fumiaki Nakata$^{2,3}$, Kazuhiro Shimasaku$^{1,4}$ and 
Sadanori Okamura$^{1,4}$}
\affil{$^{1}$Department of Astronomy, School of Science, The University of Tokyo, Tokyo 113-0033, Japan}
\affil{$^{2}$National Astronomical Observatory of Japan, Mitaka, Tokyo 181-8588, Japan}
\affil{$^{3}$Subaru Telescope, National Astronomical Observatory of Japan, 650  North A'ohoku Place, Hilo, HI 96720, USA}
\affil{$^{4}$Research Center for the Early Universe, School of Science, The University of Tokyo, Tokyo 113-0033, Japan}
\email{koyama@astron.s.u-tokyo.ac.jp}




\begin{abstract}
We present a wide-field H$\alpha$ imaging survey of the rich 
cluster CL0939+4713 (Abell\ 851) at $z=0.41$ with Suprime-Cam
on the Subaru Telescope, using the narrow-band filter NB921. 
The survey is sensitive to active galaxies with star formation
rates down to $\sim 0.3$M$_{\odot}$/yr throughout the 27$'\times$27$'$
field, and we identified 445 H$\alpha$ emitters along the large-scale
structures around the cluster. Using this sample, we find that 
(1) the fraction of H$\alpha$ emitters is a strong function of 
environment and shows a clear decline toward the cluster central 
region. We also find that (2) the color of H$\alpha$ emitters is
clearly dependent on environment. The majority of the H$\alpha$ 
emitters have blue colors with $B-I<2$, but we find H$\alpha$ 
emitters with red colors as well. Such red emitters are very rare 
in the cluster center or its immediate surrounding regions, while 
they are most frequently found in groups located far away from the 
cluster center. These groups coincide with the environment where 
a sharp transition in galaxy color distribution is seen.  
This may suggest that dusty star formation activity tends to be 
involved in galaxy truncation processes that are effective in groups,  
and it is probably related to the ``pre-processing'' that generates 
present-day cluster S0 galaxies. Finally, we confirm that (3) the 
mass-normalized integrated star formation rate in clusters 
(i.e. the total star formation rate within 
0.5$\times$R$_{200}$ from the cluster center 
divided by the cluster dynamical mass) rapidly increases with  
look-back time following approximately $\propto (1+z)^6$, and it is 
also correlated with the cluster mass.

\end{abstract}





\section{Introduction}

It has long been known that properties of galaxies are 
strongly correlated with environment where galaxies reside.
Clusters of galaxies are dominated by red early-type galaxies
with little star forming activity, while the dominant population
in the low-density field are blue late-type galaxies with 
significant star formation (e.g. \citealt{dre80}; \citealt{lew02}; 
\citealt{gom03}). It is suspected that the growth of large-scale 
structures comes into play and alters galaxy properties during the 
course of hierarchical assembly.  However, it is not yet clear what 
physical process(es) is (are) actually responsible for shaping 
galaxy properties depending on the environment.
Distant clusters of galaxies, which tend to be in an active
phase of mass assembly, are ideal sites for studying ``directly''
what is really happening along with the cluster growth.

Cosmological simulations predict that galaxies and groups 
are assembled onto massive clusters moving along filamentary 
structures (e.g. Millennium Simulation; \citealt{spr05}), 
and in fact, wide-field observations of distant clusters of galaxies 
have revealed such filamentary large-scale structures around 
rich clusters up to $z\sim 1.3$ (e.g. \citealt{kod05}; 
\citealt{tan07a}). Those cluster surrounding regions including 
in-falling groups and/or filamentary structures are likely to be 
playing critical roles in the evolution of cluster galaxies.
It is reported that the rest-frame ultra-violet(UV)--optical
colors change sharply from blue to red in such medium density environments
(e.g. \citealt{kod01b}; \citealt{tan05}; \citealt{koy08}),
suggesting that at least a part of the star forming activities
of cluster red galaxies are quenched through some environmental 
effects {\it before} entering the cluster core region. 
However, optical colors do not necessarily provide us with
the full picture of star forming activities of galaxies.
In fact, some ``red'' galaxies involve, despite of their red colors, 
a significant amount of dust-obscured  star-formation activity 
(e.g. \citealt{wol05}; 2009; \citealt{dav06}; \citealt{koy08}; 2010; 
\citealt{hai08}; \citealt{ver08}; \citealt{gal09}; \citealt{mah09}; 
\citealt{bra09}). Such galaxies might be the key population in the 
``transition phase'' from blue active galaxies to red quiescent ones. 
Therefore, it is crucial to quantify star formation activity more 
robustly using not only colors but also other independent indicators 
of star formation.

For this purpose, H$\alpha$ emission line ($\lambda _{\rm{rest}} 
=6563$\AA) is of great use as it is one of the best indicators of star 
formation which directly reflects the UV radiation from O- and B-type stars
in the \hii\ regions and is very well calibrated with local galaxies 
(\citealt{ken98}). Also, the H$\alpha$ line is much less affected by 
dust extinction or metallicity effect compared to \oii\ lines at a 
shorter wavelength ($\lambda _{\rm{rest}} =3727$\AA) or UV--optical 
colors, which are more commonly used in the studies of galaxies 
in the distant Universe.

Taking the great advantage of the wide field of view of Suprime-Cam
(\citealt{miy02}) on Subaru Telescope (\citealt{iye04}),
panoramic narrow-band imaging of H$\alpha$ emitters in the distant cluster
environment was first conducted by \cite{kod04}, who targeted the
CL0024+16 cluster at $z=0.39$ over a $\sim 27' \times 27'$ area 
centered on the cluster. Following this study, \cite{koy10} conducted 
wide-field narrow-band imaging of H$\alpha$ emitters in and around 
the RXJ1716+6708 cluster at $z=0.81$. They used the wide-field 
near-infrared camera, MOIRCS (\citealt{ichi06}; \citealt{suz08}),
on the Subaru Telescope, and spent 8 pointings to neatly cover the 
known filamentary structures found by \cite{koy07}.
\cite{fin04} and \cite{fin05} also performed H$\alpha$ emitter surveys
in the central regions of several EDisCS clusters at $z\sim$0.6$-$0.8
(EDisCS: \citealt{whi05}). More recently, \cite{sob11} studied 
environmental dependence of star forming activity at $z\sim 0.8$ based 
on their H$\alpha$ emitters sample from HiZELS (which includes several 
clusters in the COSMOS and the UKIDSS UDS fields). 
In spite of the great importance of H$\alpha$ surveys of clusters and
their surrounding regions to investigate the environmental effects,
the number of H$\alpha$ surveys of clusters has still been very limited.
Since clusters of galaxies are statistical objects by nature,
we desperately need more H$\alpha$-based studies of distant clusters
that cover a wide area and a long time baseline in order to discuss the
environmental variation of star forming galaxies and its evolution.

In this paper, we present a H$\alpha$ emitter survey for another 
cluster at $z\sim 0.4$, CL0939+4713 (Abell\ 851). This is a 
very rich cluster and is one of the most famous intermediate 
redshift clusters. Intensive imaging/spectroscopic surveys of 
this cluster including the {\it MORPHS} survey (\citealt{sma97})
have been made by many authors, which include ground-based 
broad-band imaging (e.g. \citealt{dre92}; \citealt{sta95}; 
\citealt{iye00}; \citealt{kod01b}), narrow-band imaging (\citealt{bel95}; 
\citealt{mar00}), optical spectroscopy (e.g. \citealt{dre99}; 
\citealt{sat06a}b ; \citealt{oem09}; \citealt{nak11}), 
Hubble Space Telescope imaging including UV observation (e.g. 
\citealt{dre94a}b; \citealt{sma99}; \citealt{bus00}), 
Spitzer mid-infrared imaging (\citealt{dre09}), 
sub-millimeter 850$\mu$m observation (\citealt{cow02}) and 
VLA radio observation (\citealt{sma99}). This cluster has also 
been targeted by X-ray observations several times (\citealt{sch96}
; \citealt{sch98}; \citealt{def03}).  
The X-ray image of this cluster shows two prominent peaks in X-ray
emission, which suggests that the Abell\ 851 cluster is a dynamically 
young system.  

The structure of this paper is as follows. In Section~2, we summarize 
our project and the concept of this paper. In Section~3, we show the 
selection technique of H$\alpha$\ emitters at $z=0.4$ and the derivation 
of H$\alpha$-derived star-formation rates for the selected H$\alpha$ 
emitters. Our main results and discussions are described in 
Sections~4-6, and we summarize our results in Section~7. 
Throughout this paper, we assume $\Omega_{\rm{M}} =0.3$, $\Omega_{\Lambda} =0.7$, 
and $H_0 =70$ km s$^{-1}$Mpc$^{-1}$, which gives a 1$''$ scale of 
5.41 kpc at the most up-to-date redshift of the Abell\ 851 cluster 
($z=0.405$; \citealt{oem09}). Magnitudes are all given in the AB
system. \\ \\

\section{Data}

\subsection{PISCES project}
We have been conducting the Panoramic Imaging and Spectroscopy of 
Cluster Evolution with Subaru project (PISCES: \citealt{kod05}).
We widely observed X-ray detected rich clusters at $0.4\lsim z \lsim 1.4$
mainly using Suprime-Cam on the Subaru Telescope, and discovered
prominent large-scale structures around each cluster (\citealt{kod01b}; 
\citealt{tan05}; 2006; 2007ab; 2008; 2009ab; \citealt{nak05}; 
\citealt{koy07}; 2008).   
We also conducted wide-field mapping of star formation around several 
clusters by H$\alpha$ or \oii\ line using available narrow-band filters 
installed on the Suprime-Cam or MOIRCS (\citealt{kod04};
\citealt{koy10}; \citealt{hay10}; 2011). The narrow-band imaging survey is 
powerful in the sense that it enables us to conduct a very complete
survey of star forming activity in and around clusters. 
We can detect emission lines by imaging observations and can measure 
the line flux throughout the field at the same time (see also Section~3.1). 

\subsection{This study}

\begin{figure}
\vspace{-0.7cm}
\begin{center}
 \rotatebox{0}{\includegraphics[width=9.0cm,height=9.0cm]{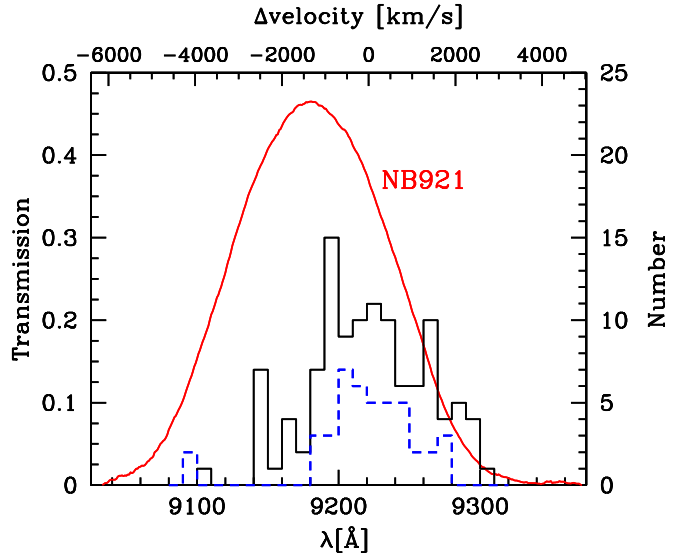}}
 \end{center}
 \vspace{-0.6cm}
\caption{ The transmission curve of the NB921 filter, including 
the response of the CCD. The histograms show the 
distribution of the line-of-sight velocity relative to the velocity center
and the corresponding observed wavelength of H$\alpha$ lines for
spectroscopically confirmed cluster member galaxies. The solid-line 
histogram is for galaxies near the cluster center and the dashed-line
histogram is for galaxies far away from the cluster, taken from Oemler et 
al.\ (2009) and Nakata et al.\ (2011), respectively. 
\label{fig:filter}}
\end{figure}
As a pioneering survey of our PISCES project, the Abell\ 851 cluster
was studied by \cite{kod01b}. They performed $BVRI$ imaging of this 
cluster with Suprime-Cam covering $27'\times 27'$ field and found 
prominent large-scale structures using the photometric redshift technique. 
\cite{nak11} performed a wide-field spectroscopic follow-up observation 
and confirmed that these huge structures are really at the 
same redshift as of the main cluster. Fortunately, the H$\alpha$ line 
from Abell\ 851 shifts to $\simeq$9220\AA {} which is between the night 
sky emission lines. We utilize a narrow-band filter NB921 ($\lambda _c 
= 9180$\AA, $\Delta \lambda = 133$\AA ) installed on the 
Suprime-Cam, and using this filter we conduct a wide-field 
H$\alpha$ emitter survey for this cluster. The transmission curve 
of the NB921 filter is shown in Fig.~\ref{fig:filter} with the 
velocity distribution of spectroscopically confirmed cluster member 
galaxies taken from \cite{oem09} (solid-line histogram). 
We also show the velocity distribution of member galaxies in the 
filamentary structures located far away from the cluster center 
taken from \cite{nak11} (dashed-line histogram). No significant 
difference can be seen between the two distributions, supporting 
that our narrow-band survey is uniformly sensitive throughout the 
field. We also note that the peak of the velocity 
distribution and that of the filter transmission are slightly different.
We can still detect H$\alpha$ emission from
cluster member galaxies with the line-of-sight velocities of
$-3500$km/s$\lsim \Delta v \lsim +1000$km/s, although
we need to be careful when deriving physical
quantities such as star formation rates (SFRs). Nevertheless, 
this filter is sensitive to $\gsim$70 \% of H$\alpha$ emission
lines of member galaxies throughout the field, and this provides
us with a great opportunity to conduct a wide-field H$\alpha$ emitter 
survey in and around the Abell\ 851 cluster. 

On top of the Suprime-Cam $BVRI$ imaging data and photometric
redshifts (phot-$z$) of the Abell\ 851 cluster field taken from \cite{kod01b}, 
we have also collected $z'$-band and NB921 data using Suprime-Cam 
covering the same area as the existing $BVRI$ imaging data 
(i.e. 27$'\times$27$'$). The data are reduced in a standard manner 
using the Suprime-Cam data reduction pipeline (\citealt{yag02}; 
\citealt{ouc04}) in the same way as described in \cite{kod01b}, 
and all the images are smoothed to $\simeq 1.1''$ seeing size, which 
is the worst seeing among our data (the $B$-band data). The exposure 
times are 30 min and 180 min for $z'$-band and NB921, respectively, 
and the 5$\sigma$ limiting magnitudes are 24.0 mag and 24.4 mag, 
respectively (measured from the deviation of randomly distributed 
3$''$ apertures in each image). Photometry of sources is performed 
at the position of $I$-band detected objects with $I\le 24.0$mag
($= M_V^* +4$) using SExtractor software (\citealt{ber96}), and 
we use 3$''$ aperture photometry throughout the paper. 

\section{Analysis}

\subsection{H$\alpha$ emitter selection}

\begin{figure}
 \begin{center}
 \vspace{-2.0cm}
 \rotatebox{0}{\includegraphics[width=9.0cm,height=9.0cm]{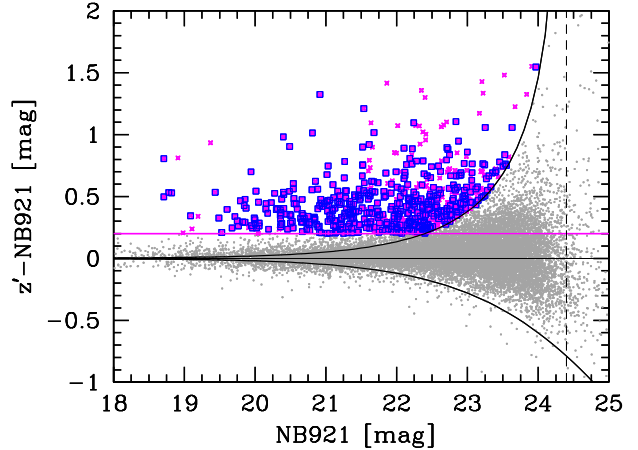}}
 \end{center}
 \vspace{-0.6cm}
\caption{ The plot showing our selection method of the NB921 emitters.
 The $z'-$NB921 colors are plotted against NB921 magnitudes.
 The vertical dashed line
 indicates the 5$\sigma$ limiting magnitude for NB921, and the
 solid-line curves show the $\pm$3$\sigma$ excesses in $z'-$NB921 
 colors. Galaxies with $z'-$NB921$>$0.2 and $z'-$NB921$>$3$\sigma$
 are defined as NB921 emitters (magenta crosses). The blue squares
 indicate the H$\alpha$ emitters at $z\sim 0.4$ selected in Fig.~3
 (i.e. NB921 emitters in the blue-line box in Fig.~3). 
\label{fig:emitter}}
\end{figure}

\begin{figure}
 \begin{center}
 \rotatebox{0}{\includegraphics[width=9.0cm,height=9.0cm]{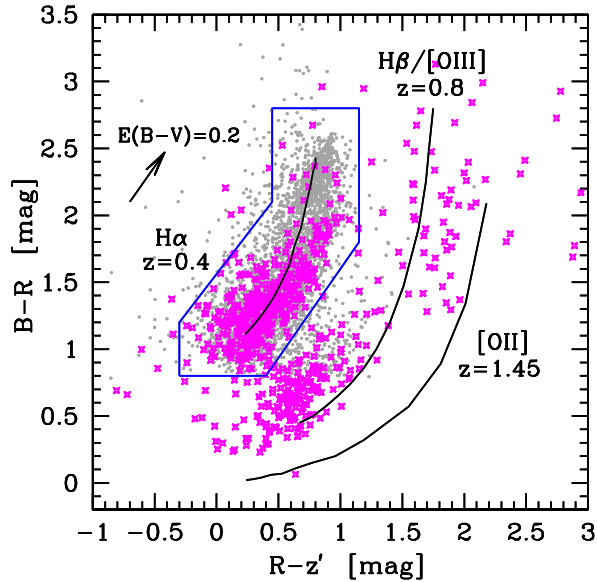}}
 \end{center}
 \vspace{-0.6cm}
\caption{ The color--color diagram to separate H$\alpha$ emitters at $z=0.4$
from other major line contaminations at other redshifts (H$\beta$/\oiii\ 
emitters at $z\sim 0.8$ and [OII] emitters at $z\sim 1.45$). Gray dots 
represent phot-$z$ selected member galaxies and magenta crosses
indicate all NB921 emitters selected in Fig.~2. Three solid-line 
curves indicate color tracks of galaxies at $z=$0.4, 0.8 and 1.45,
based on the model prediction of Kodama et al.\ (1999). Along each  
track, we change the fraction of bulge contribution to the total light
from 0.0 to 1.0 (from blue side to red side).  
We define the NB921 emitters (i.e. magenta crosses) within the blue-line 
closed box as H$\alpha$ emitters at $z=0.4$, while emitters outside the 
box as contaminant line emitters. We also show the reddening vector 
corresponding to $E(B-V)=0.2$ for $z=0.4$ galaxies, calculated from 
the extinction law of Calzetti et al.\ (2000).  
\label{fig:2color}}
\end{figure}

We first identify narrow-band excess galaxies using 
the $z'-$NB921 colors. We plot in Fig.~\ref{fig:emitter} the 
$z'-$NB921 colors of all galaxies within the observed field 
against their NB921 magnitudes. The curves show $\pm$3$\sigma$
color excesses.
We define the NB921 emitters as those satisfying
$z'-$NB921$>$0.2 and $z'-$NB921$>$3$\sigma$ (see the magenta
crosses above the $z'-$NB921$=$0.2 line and $z'-$NB921$=$+3$\sigma$ 
curve in Fig.~\ref{fig:emitter}). 
We detect 724 NB921 emitters in total, but a part of these emitters 
are not necessarily H$\alpha$ emitters at $z=0.4$ since 
H$\beta$/\oiii\ emitters at $z\sim 0.8$ and \oii\ emitters at 
$z\sim 1.45$ can also be detected as NB921 emitters.
However, these contaminations can be relatively easily eliminated
based on their broad-band colors. 
In Fig.~\ref{fig:2color}, we show the $B-R$ v.s. $R-z'$ color--color 
diagram. We here plot galaxies with $0.30<z_{phot}<0.45$ (gray dots) 
and all the NB921 emitters selected above (magenta crosses). We also 
show the model prediction of colors for galaxies at $z=$0.4, 0.8 
and 1.45 from \cite{kod99}. This color combination clearly separates 
the H$\alpha$ emitters at $z\sim 0.4$ from other major line emitters at 
other redshifts, and in fact, these colors are also used in 
\cite{kod04} in order to identify H$\alpha$ emitters at a similar 
redshift ($z=0.39$). We define the NB921 emitters distributed 
in the closed box in Fig.~\ref{fig:2color} as H$\alpha$ emitters 
at $z=0.4$. We note that this boundary is set by eye, and we may miss a 
small fraction of real members near the boundary. However, a small
change of this boundary does not change our results at all
(see similar selection method in \citealt{kod04} and \citealt{koy10}). 
Also, this color selection can be applied only for the galaxies
that are detected in all $BRz'$-bands. Therefore, we do not include any
H$\alpha$ emitter if it is not detected in any of these bands.
However, the quantitative analyses presented in this paper are 
mainly based on bright galaxies with $z'<23$ mag, and are not 
affected by the faint undetected objects.
After application of all these color cuts, we are left with 
445 H$\alpha$ emitters in total in and around 
the Abell\ 851 cluster throughout the observed field. 
This is one of the largest samples of H$\alpha$ 
emitters currently available for distant clusters. 

\subsection{Necessity of narrow-band imaging}

\begin{figure}
\begin{center}
 \rotatebox{0}{\includegraphics[width=9.0cm,height=9.0cm]{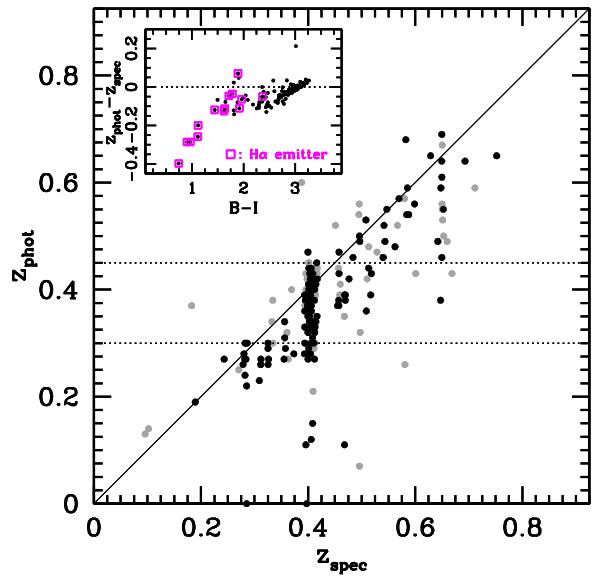}}
 \end{center}
\vspace{-0.6cm}
\caption{ The $z_{spec}$ v.s. $z_{phot}$ plot for galaxies in the
Abell\ 851 field. Spectroscopic data are taken from Table 1 of Oemler et
 al.\ (2009). Black and gray symbols indicate the galaxies with 
Q$=$1,2 and Q$=$3,4, respectively, where Q means the quality of the
spectra (smaller Q means better quality; Oemler et al.\ 2009). 
The horizontal dotted lines show the phot-$z$ member selection 
criteria used in this paper. 
In the inset, we show the accuracies of our photometric redshifts 
for spectroscopic members (i.e. $0.38\le z_{spec} \le 0.42$) as a 
function of $B-I$ color. Open squares indicate H$\alpha$ emitters
selected in Section~3.1. Photometric redshifts sometimes return lower
values for blue galaxies, but it is evident that they are neatly 
recovered as the H$\alpha$ emitters. 
\label{fig:specz_photz}}
\end{figure}

We here briefly describe the powerfulness of the narrow-band survey. 
We show a plot showing $z_{spec}$ v.s. $z_{phot}$
in Fig.~\ref{fig:specz_photz}. The spectroscopic catalog of the 
Abell\ 851 field is taken from \cite{oem09}.  
The photometric redshifts are not always perfect, especially for blue 
star forming galaxies due to the lack of significant features such 
as 4000\AA\ break in their spectral energy distribution 
(SED) (e.g. \citealt{kod99}). In fact, as can be seen in 
Fig.~\ref{fig:specz_photz}, our photometric redshifts for blue member
galaxies sometimes fail and return significantly lower redshifts
(see the inset of Fig.~\ref{fig:specz_photz}). This color dependence 
of the phot-$z$ accuracy makes it difficult to study star 
forming activity and/or its dependence on environment. 
In contrast, narrow-band surveys detect emission lines as an excess
of flux at a certain wavelength, and in fact, many of the real 
members missed in the phot-$z$ selection can be {\it rescued} 
in our narrow-band survey (see the squares in the inset of 
Fig.~\ref{fig:specz_photz}). Thus, we again stress that a more complete 
sample of star-forming galaxies can be constructed regardless of their
colors, through the narrow-band survey (see also the discussion in 
\citealt{kod04}).   

\subsection{H$\alpha$-derived star formation rate}

Then, we estimate the star formation rates (SFR) for all the H$\alpha$
emitters selected above. At first, we calculate the H$\alpha$+\nii\ line
flux ($F_{\rm{H}\alpha + \rm{[NII]}}$), continuum flux density ($f_c$) 
and rest-frame equivalent width (EW$_{R}$) in the following equations:
\begin{equation}
F_{\rm{H\alpha + [NII]}}=\Delta_{\rm{NB}}\frac{ f_{\rm{NB}} -
 f_{z'} }{1-\Delta_{\rm{NB}}/\Delta_{z'}}
\end{equation}
\begin{equation}
f_c = \frac{f_{z'} - f_{\rm{NB}}(\Delta_{\rm{NB}}/\Delta_{z'})}{1-\Delta_{\rm{NB}}/\Delta_{z'}}
\end{equation}
\begin{equation}
\textrm{EW}_R({\rm{H\alpha + [NII]}}) = (1+z)^{-1} \frac{F_{\rm{H\alpha}
 + [NII]}}{f_c} 
\end{equation}
where $\Delta_{z'}$ ($=955$\AA {}) and $\Delta_{\rm{NB}}$ ($=133$\AA )
are the FWHMs of the $z'$ and NB921 filters, $f_{z'}$ 
and $f_{\rm{NB}}$ are the flux densities at $z'$-band and at NB921, 
respectively. We then multiply 
4$\pi$$d_L^2$ by $F_{\rm{H\alpha +[NII]}}$ to derive the luminosity 
$L$(H$\alpha$+ [NII]), where $d_L$ is the luminosity 
distance of 2204 Mpc at $z=0.405$. 

Finally, we compute the H$\alpha$-based star formation rates, 
SFR(H$\alpha$), using the Kennicutt (1998) relation for Salpeter (1955)
IMF: SFR(H$\alpha$)[$M_{\odot}$/yr]$=7.9\times 10^{-42}
L_{\rm{H\alpha}}$[erg/s]. We correct for 30\% \nii\ line contribution 
(e.g. \citealt{tre99}) and consider 1 mag extinction in H$\alpha$ flux
due to dust (e.g. \citealt{ken83}; \citealt{ken92}),
following our previous studies (\citealt{kod04}; \citealt{koy10}).  
Our selection criteria of H$\alpha$\ emitters shown in Section~3.1 
correspond to EW$_{\rm{R}}$(H$\alpha$+\nii)
$\gsim$20\AA\ and SFR(H$\alpha$)$\gsim$0.3$M_{\odot}$/yr, 
and the typical uncertainty in the SFR(H$\alpha$) from
photometric error is $\sim 0.1 M_{\odot}$/yr. 
We should note that the assumptions for \nii\ contamination
and dust extinction adopted above are somewhat 
uncertain. It is reported that the contribution of \nii\ emission to 
the H$\alpha$+\nii\ line flux depends on EW(H$\alpha$+\nii) 
(e.g. \citealt{vil08}) or B-band luminosity (\citealt{ken08}). 
Also, the strength of dust extinction can vary a lot
(e.g. depending on the morphological type of galaxies: \citealt{bos01}), 
and in the extreme cases, it may become A(H$\alpha$)$\gsim 3$mag 
(e.g. \citealt{pog00}), although the assumption of $\sim$1 mag
extinction adopted here seems to be valid on average in the 
distant Universe as well and the level of dust extinction
at a given SFR does not change strongly with redshift 
(e.g. \citealt{gar10}; \citealt{moo10}). \\

\begin{figure*}
 \begin{center}
 \rotatebox{0}{\includegraphics[width=17.0cm,height=17.0cm]{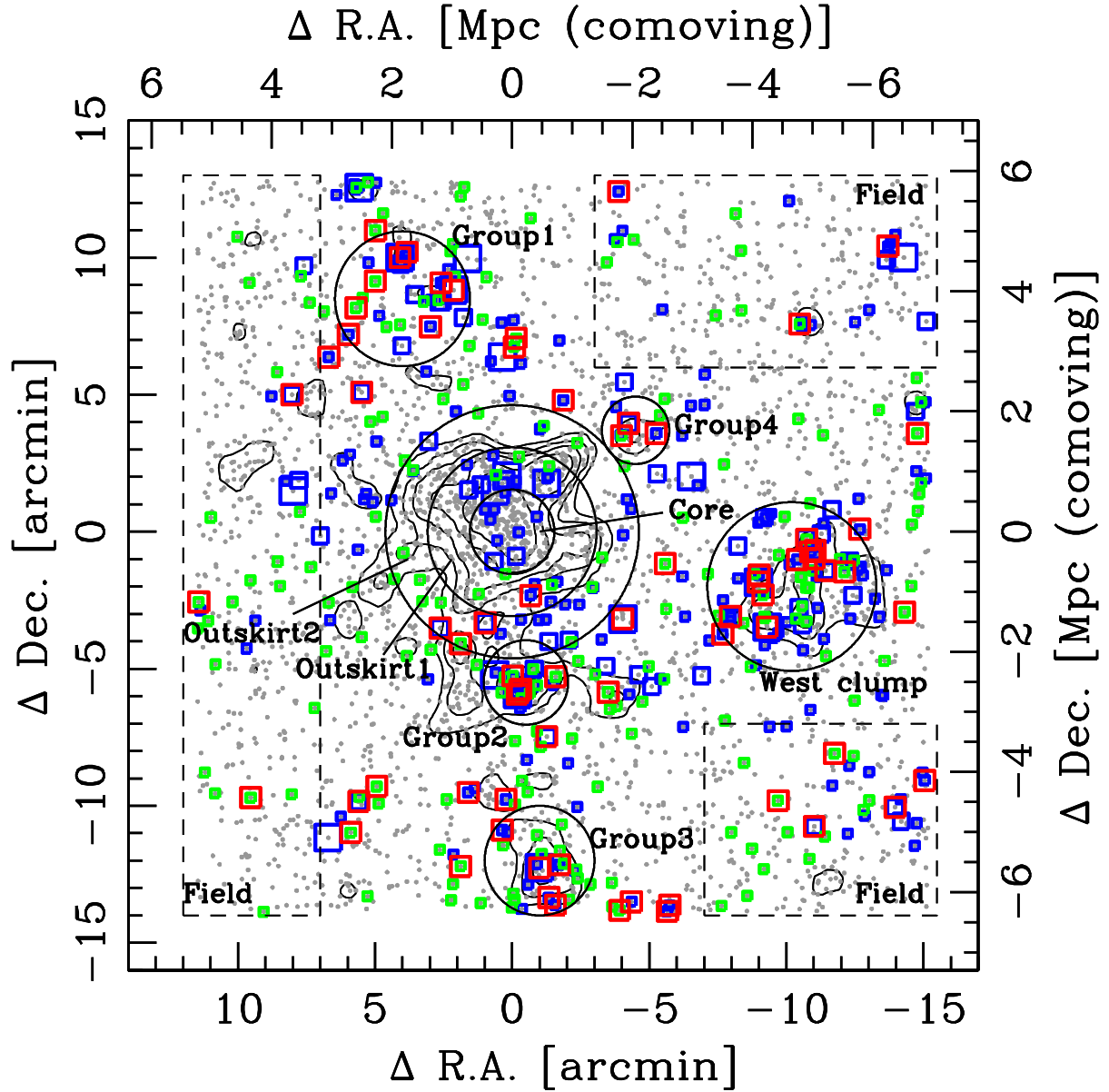}}
 \end{center}
\vspace{-0.6cm}
\caption{ Spatial distribution of the phot-$z$ members with
 0.30$<z_{phot}<$0.45 (gray dots) and H$\alpha$ emitters 
(open colored squares). 
 Blue squares indicate H$\alpha$ emitters with SFR$_{\rm{H\alpha}}$$>$0.75
 M$_\odot$/yr and larger symbols indicate galaxies with larger SFRs. 
 Green squares show weak H$\alpha$ emitters with 
 SFR$_{\rm{H\alpha}}$$<$0.75M$_{\odot}$/yr. Red squares represent the red 
 H$\alpha$ emitters with $B-I>2.0$. Contours show 3,5,7,10, and 15$\sigma$ 
 significance of galaxy overdensity calculated using all the member 
 galaxies (all the phot-$z$ members and/or H$\alpha$ emitters). 
 Solid-line circles and dashed-line rectangles show the areas where we define
 different environments to study environmental effects (see text and 
 Table~1). 
\label{fig:wide_map}}
\end{figure*}

\section{Wide-field mapping of H$\alpha$ emitters}

\subsection{Panoramic H$\alpha$ view of the Abell\ 851 cluster }

\begin{table*}
\begin{center}
\caption{Definition of environment around Abell\ 851. \label{tbl-1}}
\begin{tabular}{p{2.3cm}p{1.9cm}p{1.9cm}p{1.6cm}p{1.6cm}p{1.6cm}}
\hline 
\hline
     & R.A. & Dec. & $\Delta$R.A. & $\Delta$Dec. & radius    \\
Name & (J2000) & (J2000) & (arcmin) & (arcmin) & (Mpc) \\
\hline 
Core................ &  09 42 58.0  &  +46 59 01  & 0.0  & 0.0 &    0.5    \\
Outskirt-1.......... &  09 42 58.0  &  +46 59 01  & 0.0  & 0.0 & 0.5$-$1.0    \\
Outskirt-2.......... &  09 42 58.0  &  +46 59 01  & 0.0  & 0.0 & 1.0$-$1.5    \\
West Clump......     &  09 41 57.7  &  +46 56 56  & $-$10.2 & $-$2.0 & 1.0  \\
Group-1............. &  09 43 21.5  &  +47 07 37  & 4.0 & 8.5 & 0.8  \\
Group-2............. &  09 42 55.2  &  +46 53 28  & $-$0.5 & $-$5.5  & 0.5  \\
Group-3............. &  09 42 52.4  &  +46 46 54  & $-$1.0 & $-$12.0 & 0.65 \\
Group-4............. &  09 42 31.2  &  +47 02 44  & $-4.5$ & 3.7     & 0.4 \\
\hline
\end{tabular}
\vspace{3mm}
\tablecomments{$\Delta$R.A. and $\Delta$Dec. are the central coordinates 
of each environment relative to the cluster center. }
\end{center}
\end{table*}

We show in Fig.~\ref{fig:wide_map} the spatial distribution of the 
H$\alpha$ emitters selected in the previous section. 
The coordinates are shown with respect to the peak of the diffuse
X-ray emission from the intra-cluster medium
(R.A.$=$09$^h$42$^m$58$^s$.0 and Dec.$=$+46$^{\circ}$59$'$01$''$), 
which is the same definition as in \cite{oem09}.  
We plot all the cluster member candidates ($0.30 \le z_{phot} \le 0.45$; 
see the horizontal dotted-lines in Fig.~\ref{fig:specz_photz}) with 
gray dots. The H$\alpha$ emitters are shown by colored symbols in 
Fig.~\ref{fig:wide_map}. Larger blue squares indicate 
larger star formation rates, and green squares indicate the galaxies
with weak H$\alpha$ lines (SFR$_{\rm{H\alpha}}$ $<$ 0.75M$_{\odot}$/yr). 
Also, we marked red squares for H$\alpha$ emitters with $B-I>2.0$
(hereafter red H$\alpha$ emitters).   
Our survey revealed the entire distribution of star forming 
galaxies around the Abell\ 851 cluster across the $\sim$30$'$ field 
for the first time. 
It is evident that the H$\alpha$ emitters are widely distributed 
throughout the field, which strongly supports the existence of 
real structures at $z=0.41$ associated with the central cluster 
Abell\ 851. In fact, most of the structures have been spectroscopically 
confirmed to be physically associated to the cluster based on our
intensive spectroscopic follow-up (\citealt{nak11}). 

We here define the galaxy environment used in this paper as shown 
in Fig.~\ref{fig:wide_map}, and summarize them in Table~1. 
The cluster core and outskirt regions (1 and 2) correspond
to $r_c<$ 0.5 Mpc (which include the two X-ray emission
peaks; see \citealt{oem09}), 0.5 $<r_c<$ 1.0 Mpc and 1.0 $<r_c<$ 1.5 
Mpc from the cluster center, respectively (all in the physical scale). 
We also define the West clump (the richest group in the observed 
field) and Group-1,2,3, and 4 to pick out galaxies concentrated in
groups relatively far out from the cluster core. 
Note that the West clump, Group-2 and 3 are large clumps 
spectroscopically confirmed by \cite{nak11}, while the Group-1 
region was not well covered by their spectroscopic observation. 
The Group-4 is also a spectroscopically confirmed 
in-falling group (which is identical to the NW group in \citealt{oem09}). 
It is interesting to note that in addition to these known structures 
(West clump, Group-2,3,4) H$\alpha$ emitters are concentrated in the 
northern part (i.e. Group-1), where we did not identify a prominent 
overdensity of phot-$z$ selected galaxies in \cite{kod01b}. 
Therefore, our wide-field emission-line survey suggests the existence 
of a prominent structure traced by star-forming galaxies in the 
north direction of Abell\ 851. This new group (Group-1) is defined 
as a circle in Fig.~\ref{fig:wide_map}. Finally, as a comparison, 
we define the three ``Field'' regions as indicated in 
Fig.~\ref{fig:wide_map}, in which we avoid any prominent structures.

\subsection{H$\alpha$ fraction}

\begin{figure}
 \vspace{-7mm}
 \begin{center}
 \rotatebox{0}{\includegraphics[width=9.4cm,height=9.4cm]{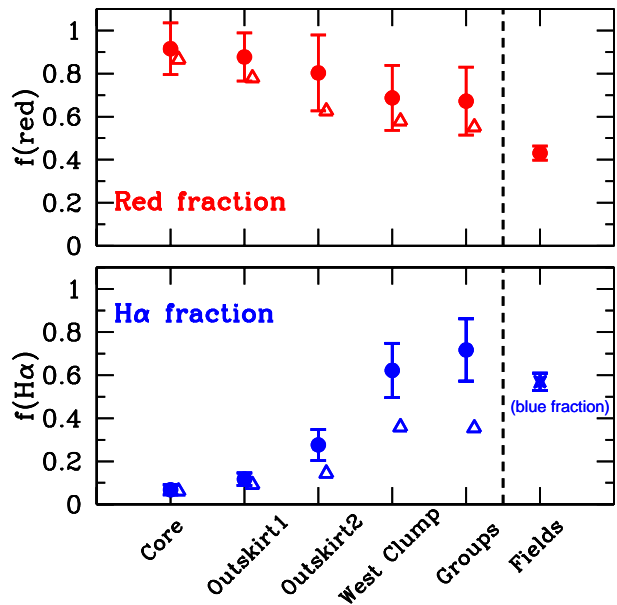}}
 \end{center}
 \vspace{-0.6cm}
\caption{ The H$\alpha$ fraction (bottom) and the red galaxy fraction (top)
as a function of environment, calculated for the galaxies with $z'<$ 23 mag.
The open triangles and filled circles indicate the values before and 
after the statistical subtraction of the contaminant galaxies, 
respectively. The error bars present the Poissonian errors. For the 
field environment, we show only the red galaxy fraction before
statistical correction and the blue galaxy fraction instead of 
H$\alpha$ fraction (see text).
\label{fig:Ha_fraction}}
\end{figure}

We here calculate the fraction of H$\alpha$ emitters for each 
environment defined above (see the labels in 
Fig.~\ref{fig:Ha_fraction}). We statistically subtract the 
expected contaminations in our phot-$z$ based membership 
using the surface number density of the ``Field'' regions, $\rho_{cont} =$ 
2.06 arcmin$^{-2}$. We calculate the H$\alpha$ fraction 
in each environment using the following equation:
\begin{equation}
f(\textrm{H}\alpha) = \frac{N_{H\alpha}}{N_{member} - N_{cont} }, 
\end{equation}
where $N_{H\alpha}$, $N_{member}$ and $N_{cont}$ represent the numbers
of H$\alpha$ emitters, all member galaxies (phot-$z$ members 
or H$\alpha$ emitters) and the expected contaminations (i.e. 
$N_{cont} = \rho_{cont} \times S$, where $S$ is the surface area of 
each environment), respectively. 
In this calculation, we limit our sample to the galaxies with
$z' \le 23$ mag, since the selection of H$\alpha$ emitters becomes 
incomplete below this magnitude as can be seen in Fig.~3. 
In Fig.~\ref{fig:Ha_fraction}, filled circles with error bars indicate 
the values after the statistical subtraction, while open triangles show
those before subtraction. It can be seen that the H$\alpha$ fractions 
are almost unchanged after statistical subtraction in the dense
cluster central region, while the contribution of 
contaminant galaxies gets larger in poorer environments. 
Note that, since we can not calculate the H$\alpha$ fraction 
for the field environment by definition (see equation 4),
we present instead the blue galaxy fraction calculated based on the
phot-$z$ selected galaxies in the field environment as a rough 
estimate of the fraction of star forming galaxies. 
This assumption would be reasonable because we find
that the fraction of H$\alpha$ emitter among blue galaxies
is $\sim 1$ at all environments after statistical correction. 
However, as will be discussed in the next section, there must be 
H$\alpha$ emitters with red colors as well.
Therefore, the blue galaxy fraction shown here for field environment 
may be a lower limit of the H$\alpha$ fraction. 
We also note that, although our ``Field'' environment avoids any
prominent structures, it is still located near the rich cluster. 
This may lead to an overestimation of contaminant galaxies in the above
calculation. However, our results including the ``relative'' trends
seen in Fig.~\ref{fig:Ha_fraction} would not change even if we do not 
apply any correction for the contamination. 

In Fig.~\ref{fig:Ha_fraction}, it is evident that the H$\alpha$ 
emitter fraction sharply declines toward the higher-density 
environment (only $\sim$10\% in the core and Outskirt-1, in 
contrast to $\gsim$50\% in the west clump and groups), 
suggesting the clear environmental dependence of star forming 
activity at $z\sim 0.4$.  We also show the fraction of 
red galaxies with $B-I>2$ in the top panel of 
Fig.~\ref{fig:Ha_fraction} (again, open triangles and filled 
circles indicate before and after statistical subtraction). 
The trend of increasing fraction of red galaxies toward the 
cluster center is consistent with our result obtained from 
the analysis on the H$\alpha$ fraction (i.e. the bottom 
panel of Fig.~\ref{fig:Ha_fraction}). 

It is now clear that the fraction of star forming galaxies
is significantly lower in the cluster central regions than in
the other surrounding environments (see also \citealt{sat06b} for 
the measurements of \oii\ emitter fraction within $\sim$3 Mpc from 
the cluster), although some H$\alpha$ emitters are found 
in the core of Abell\ 851 (see Fig.~\ref{fig:wide_map} and 
\citealt{oem09}). Such H$\alpha$ emitters found in the cluster 
central region are not distributed uniformly over the cluster 
environment but concentrated along the north-south direction near 
the core. In particular, we find a strong concentration of 
H$\alpha$ emitters at $\sim 1.5'$ north from the cluster center, 
and this position coincides with the ``North group'' 
noted by \cite{oem09} where they also found a large number of star 
forming galaxies. Therefore, it may be possible that a part of 
the star forming galaxies found in the cluster central region 
belong to a group moving near the cluster central region in projection.

\section{Red Star Forming Galaxies}

We have mapped out the spatial distribution of the H$\alpha$ emitters 
across the $\sim 30'$ field around Abell\ 851. The majority of the H$\alpha$ 
emitters have blue colors with $B-I\le 2$. This is a natural consequence 
of young stellar populations in actively star forming galaxies. On the 
other hand, we have also identified a large number of red H$\alpha$
emitters throughout the field (red squares in Fig.~\ref{fig:wide_map}). 
We here focus on the environment and nature of such red H$\alpha$
emitters, because they may be transitional galaxies migrating from 
the blue cloud to the red sequence. 

\subsection{Environment of red H$\alpha$ emitters}

\begin{figure}
 \begin{center}
 \rotatebox{0}{\includegraphics[width=9.6cm,height=9.6cm]{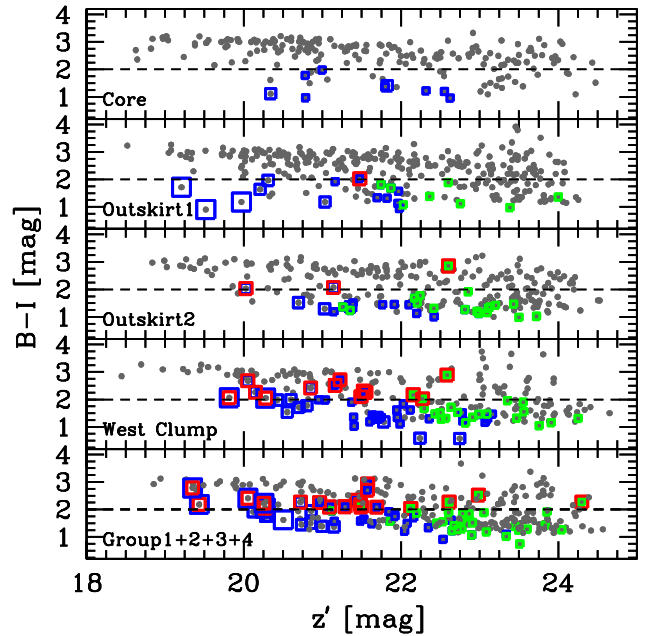}}
 \end{center}
 \vspace{-0.6cm}
\caption{ The color--magnitude diagram for each environment as indicated in
each panel. The definition of the environment is shown in 
Fig.~\ref{fig:wide_map} and the meanings of the symbols are also the 
same as in Fig.~\ref{fig:wide_map}. The horizontal
dashed line at $B-I=2$ is the dividing line between red and blue 
galaxies in this paper. It is clear that the red H$\alpha$ emitters are 
much more common in the group-scale environments.  
\label{fig:colmag}}
\end{figure}

\begin{figure}
\vspace{-1.5cm}
 \begin{center}
 \rotatebox{0}{\includegraphics[width=9.0cm,height=9.0cm]{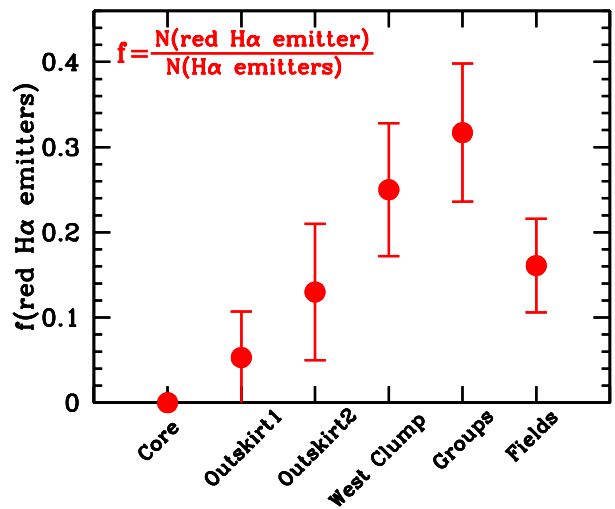}}
 \end{center}
\vspace{-0.6cm}
\caption{ The fraction of red galaxies among all H$\alpha$ emitters 
with $z'<23$ mag. A clear deficit of red H$\alpha$ emitters 
is seen in the cluster central regions. Note that only the H$\alpha$
emitters are used in this calculation, so that no statistical 
subtraction is needed. The error bars indicate Poissonian uncertainties.
\label{fig:red_fraction}}
\end{figure}

To examine the colors of the H$\alpha$ emitters and its environmental 
dependence in detail, we construct color--magnitude diagrams 
in Fig.~\ref{fig:colmag}. We divide the sample into five environmental bins 
as defined in the previous section, namely, cluster core, two cluster 
outskirts, west clump and groups (Group-1,2,3, and 4). 
We use the same color symbols as used in Fig.~\ref{fig:wide_map} 
(i.e. blue, green and red squares for normal, weak and 
red H$\alpha$ emitters, respectively). 
Fig.~\ref{fig:colmag} shows the most important and impressive
result of this paper.  We find almost no red H$\alpha$ emitters 
in the central cluster regions (i.e. in the cluster core and the two 
outskirt regions). In contrast, we see a large number of red emitters 
in the west clump and in the groups.
To quantify this, we calculate the fraction of red galaxies among 
all H$\alpha$ emitters in each environment (see Fig.~\ref{fig:red_fraction}). 
This plot clearly shows that the red H$\alpha$ emitters are seen exclusively
in the relatively low-density environments.
We find that a surprisingly high fraction
($\gsim$20--30\%) of H$\alpha$ emitters in the west clump or in the groups 
show red colors, which probably suggests that dusty star
forming activities are triggered in such environments (see also
discussion in Section~5.3). 


Note that the definition of ``red galaxies'' here includes not 
only the galaxies on the red-sequence but also those in the 
``green-valley''. However, our result is not affected even if we 
select only the red-sequence galaxies, although the number of 
red H$\alpha$ emitters becomes smaller in this case. 
We combined the four groups to obtain a composite value for
the group environment so that we can achieve better statistics, 
although we find that the fraction is nearly constant at 
$\sim 30$\% in all the groups. 
It is also interesting to note that we find three red H$\alpha$ emitters
in the Outskirt-2 region (see the middle panel of Fig.~\ref{fig:colmag}),
but they are all located in the southern filament connecting the cluster 
core and the Group-2 (see Fig.~\ref{fig:wide_map}). This may suggest that
the galaxies falling into the cluster along the filamentary structure 
are experiencing somewhat different environmental effects from those 
for galaxies falling directly into the cluster from other various directions.

It is clear that the red H$\alpha$ emitters are absent 
in the cluster central region. We note that this deficit of 
red H$\alpha$ emitters does not simply reflect the environmental 
dependence of the color distribution of overall galaxy populations.
As shown in the top panel of Fig.~\ref{fig:Ha_fraction}, 
the fraction of red galaxies monotonically increases toward the 
cluster center (equally, those H$\alpha$ emitters near the 
cluster core are almost exclusively blue).
Therefore, the environmental trend of the red galaxies
and that of the red H$\alpha$ emitters are clearly different, and so
there is a clear difference in star forming activity 
between cluster and group/filament environments.

One may claim that the definitions of the groups and the west 
clump are somewhat arbitrarily chosen.
In order to assess the effect of the uncertainty in the
definition of environment, we apply another definition of environments, 
based on the local density of galaxies (i.e.\ $\Sigma _{10th}$ 
using all the phot-$z$ members). We then investigate the dependence
of colors of H$\alpha$ emitters on the local density. The trends seen in 
Figs.~\ref{fig:colmag} and \ref{fig:red_fraction} are still visible, 
but weaker. In fact, we find that the local densities 
of the Outskirt-2 and the groups are similar, but the occurrence
of the red H$\alpha$ emitters is different ($\sim$ twice larger 
in the groups than in the Outskirt-2, although the error-bars are large). 
This may suggest that star forming activity is determined not 
solely by local environment but also by global environment. 

We should note that the preferred environment of the red H$\alpha$ emitters
may change with redshift. In fact, we have shown in \cite{koy10} that
red H$\alpha$ emitters
are located immediately outside of the cluster core at $z\sim 0.8$. 
However, our current analysis on Abell\ 851 suggests that at $z\sim 0.4$ 
red H$\alpha$ emitters are found in group environment relatively far away 
from the cluster, and that such galaxies are very rare even in the 
cluster outskirts. This may support the ``propagation scenario'' 
of star formation in clusters, that is, the site of the
red star forming galaxies (probably the transition objects) shifts from
cluster cores to outer regions from $z\sim 1.5$ to 
the present-day Universe (see also \citealt{hay10}).  However,
the situation might be more complicated, given that there exist
heavily obscured mid-infrared (MIR) bright galaxies as well. 
A more detailed discussion including such MIR sources will 
follow in the next sections. \\ 

\subsection{Mid-infrared sources}

\begin{figure}
 \begin{center}
 \rotatebox{0}{\includegraphics[width=9.0cm,height=9.0cm]{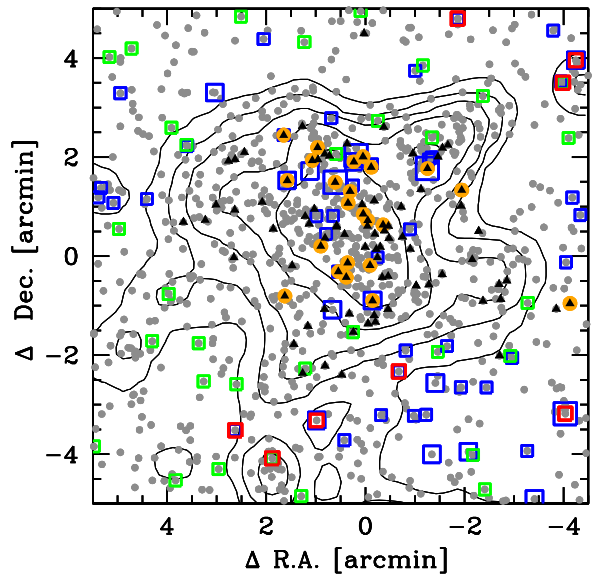}}
 \end{center}
 \vspace{-0.6cm}
\caption{ A close-up view of the cluster central region with the positions
of MIPS-detected sources (orange circles). The meanings of the other symbols
are the same as in Fig.~5, except for the black triangles which show the 
spectroscopic members within the MIPS data coverage, taken from Oemler
et al.\ (2009). \label{fig:map_center}}
\end{figure}

\begin{figure}
\vspace{-2.5cm}
 \begin{center}
 \rotatebox{0}{\includegraphics[width=9.0cm,height=9.0cm]{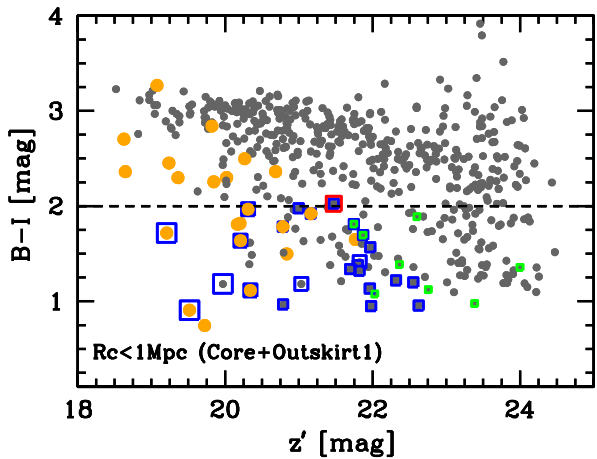}}
 \end{center}
 \vspace{-0.6cm}
\caption{ A color--magnitude diagram for the cluster central region
 ($R_c < 1$Mpc). The meanings of the symbols are the same as in Fig.~9.
MIPS sources are again shown by orange filled circles. 
\label{fig:colmag_center}}
\end{figure}

The central $\sim 5'\times 5'$ region of Abell\ 851 was observed with 
Spitzer MIPS(24$\mu$m) by \cite{dre09}. They reported that 
some galaxies in the core of Abell\ 851 are detected in MIR.
Here, we show in Fig.~\ref{fig:map_center} the spatial distribution 
of such MIPS sources taken from the catalog in \cite{oem09} overlaid 
on the distribution of our H$\alpha$ emitters. We only show the 
cluster central region due to the small coverage of the MIPS data.
The MIPS sources are shown by orange circles.  
It is apparent that the spatial distribution of the H$\alpha$ emitters and 
the MIPS sources are qualitatively similar, and they are both concentrated
in the north-south direction. In fact, some sources are directly 
overlapping each other. We also construct a color--magnitude diagram 
for the central region with $R_c < $1 Mpc (Fig.~\ref{fig:colmag_center}). 
The MIPS sources are again shown by orange circles 
in Fig.~\ref{fig:colmag_center}. They are located on the slightly 
bluer side of the red sequence and also at the bright end of the
``blue cloud'' and the ``green valley''. 
Some of the MIR galaxies are overlapping with 
H$\alpha$ emitters, but interestingly, most of the red MIR galaxies 
with $B-I>2$ are {\it not} detected in H$\alpha$, although these 
MIPS sources are all spectroscopically confirmed members and many of 
them should have been detected as H$\alpha$ emitters judging
from their spectroscopic redshifts (see below).
Therefore, we should keep in mind that the lack of red H$\alpha$
emitters in the cluster central region does not necessarily 
mean the lack of ``red star forming'' galaxies there. 

\begin{table*}
\begin{center}
\caption{Properties of spectroscopically confirmed MIPS sources in Abell\ 851. \label{tbl-2}}
\begin{tabular}{lcrcccrc}
\hline 
\hline
  &  & $B-I$ & EW$_R$(H$\alpha$+[NII]) & SFR(H$\alpha$) & $f_{24\mu m}$ &
 SFR(IR) & H$\alpha$ emitter \\
redshift & spectral type & [mag] & [\AA] &
 [M$_\odot$/yr] & [Jy] & [M$_\odot$/yr] & (yes/no) \\
\hline 
0.4084 & e(c) & 1.11 &  34.52 &  3.51 & 3.05 $\times$10$^{-4}$ &  10.2 & yes \\
0.4059 & e(a) & 1.78 &  29.04 &  1.98 & 1.63 $\times$10$^{-4}$ &   5.0 & yes \\
0.3972 & e(n) & 0.75 & 147.31 & 22.93 & 7.96 $\times$10$^{-4}$ &  30.1 & yes \\
0.4010 & e(a) & 1.92 &  30.84 &  1.48 & 1.30 $\times$10$^{-4}$ &   3.8 & yes \\
0.3958 & e(b) & 0.91 & 145.75 & 27.45 & 1.04 $\times$10$^{-3}$ &  40.3 & yes \\
0.3932 & e(a) & 1.64 &  28.90 &  3.35 & 2.12 $\times$10$^{-4}$ &   6.8 & yes  \\
0.4061 & e(n) & 1.72 &  69.79 & 19.24 & 3.98 $\times$10$^{-3}$ & 165.5 & yes \\ 
0.3937 & e(a) & 1.97 &  38.64 &  4.02 & 4.82 $\times$10$^{-4}$ &  17.2 & yes \\
\hline
0.4007 & k+a  & 2.70 &  8.66  &  4.44  & 5.21$\times$10$^{-4}$ &  18.7 & no \\  
0.4060 & k+a  & 2.36 &  3.59  &  1.82  & 3.14$\times$10$^{-4}$ &  10.6 & no \\
0.4075 & a+k  & 2.30 & $-$2.57 & $-$0.37 & 1.12$\times$10$^{-4}$ & 3.2 & no \\
0.4076 & e(a) & 1.65 & 21.00  &  0.59  & 8.13$\times$10$^{-5}$ &   2.2  & no \\
0.3938 & k+a  & 2.30 & 14.27  &  3.69  & 5.78$\times$10$^{-4}$ &  21.1 & no  \\
0.4083 & k+a  & 2.50 &  0.91  &  0.10  & 1.49$\times$10$^{-4}$ &   4.5 & no \\
0.3960 & k    & 2.84 &  0.40  &  0.07  & 7.32$\times$10$^{-4}$ &  27.4 & no  \\
0.4017 & k    & 3.17 & $-$0.50 & $-$0.04 & 1.01$\times$10$^{-4}$ & 2.9 & no  \\
\hline 
\end{tabular}

\vspace{3mm}
\tablecomments{ All the galaxies listed here are spectroscopically 
confirmed MIPS-detected members at $0.389 \le z_{spec} \le 0.409$
(i.e. H$\alpha$ lines of these sources fall within the FWHM of the
NB921 filter), with spectral quality Q$\le 3$ in Oemler et al. (2009). 
Remarkably, the MIPS sources without H$\alpha$ emissions (lower side)
tend to be redder and to have spectral types without emission 
lines (i.e. k/k+a/a+k) compared to those with H$\alpha$ detections 
(upper side). Redshifts, spectral types and $f_{24\mu m}$ are taken from 
Oemler et al. (2009).  In deriving SFR(IR), we first estimate
$L$(IR) by using the SED templates of starburst galaxies in 
Lagache et al.\ (2004) and then convert it to SFR(IR) based on the 
Kennicutt (1998) calibration. Two sources have negative values of 
EW(H$\alpha$+[NII]) and SFR(H$\alpha$) due to their negative 
(but almost $\sim$0) $z'-$NB921 colors. }
\end{center}
\end{table*}

\cite{dre09} examined the spectral type of MIPS-detected sources and
found that many of the MIPS sources in Abell\ 851 have ``e(a)-type'' spectra
(strong Balmer absorption {\it and} \oii\ emission lines), which are
interpreted as on-going dusty starbursts (e.g. \citealt{pog99};
\citealt{pog00}). \cite{dre09} also noted that a significant fraction 
($\sim$30\%) of ``k+a'' or ``a+k'' galaxies (strong Balmer absorption 
{\it without} \oii\ emission lines), which are often interpreted as
{\it post-starburst} galaxies, are detected in MIR 
(see also \citealt{sma99} for the radio continuum detection 
with VLA from k+a galaxies in the Abell\ 851). 
In the spectroscopic catalog of Abell\ 851 in \cite{oem09}, 
22 galaxies with Q$\le$3 are detected with MIPS (i.e. $f_{24\mu m} 
\ge 80\mu$Jy or SFR$_{\rm{IR}} \gsim 3$M$_{\odot}$/yr). 
We find that 6 out of these 22 sources have redshifts slightly 
outside of our H$\alpha$ survey (their H$\alpha$ lines fall outside of the 
FWHM of the NB filter), but we can expect to detect H$\alpha$ 
emission from the remaining 16 sources. 
However, we find that only 8 galaxies (50\%) satisfy our 
H$\alpha$ emitter selection criteria, while the remaining 8 
galaxies (50\%) are not detected in H$\alpha$ (see Table 2).
These galaxies are likely to be heavily attenuated by dust, so that
even the H$\alpha$ emission cannot come through. Furthermore, we 
find a systematic difference in spectral types between H$\alpha$-detected and 
H$\alpha$-undetected MIPS sources (again, see Table 2). 
All the H$\alpha$-detected MIPS sources 
show ``e(a,b,c,n)'' spectra (i.e. \oii\ emission lines are present 
in their spectra), while many of the H$\alpha$-undetected MIPS sources 
show ``k+a/a+k'' spectra (or ``k''-type in two cases) 
without \oii\ emission lines. We can also confirm in Table 2 that 
the H$\alpha$-undetected MIPS sources have systematically redder 
colors than the H$\alpha$-detected ones 
($\leftangle$$B$$-$$I$$\rightangle = 1.48$ 
and $2.48$ for H$\alpha$-detected and H$\alpha$-undetected objects,
respectively). These results suggest that some red galaxies do not 
show H$\alpha$ or \oii\ emission lines in spite of their significant 
amount of hidden star formation activities.

\subsection{What are the red H$\alpha$ emitters ?}

It is naturally expected that star forming galaxies show blue colors
because young, massive, hot stars dominate their total light.
Therefore, the interpretation for the red galaxies with emission lines
is not straightforward. The most likely interpretation for the red 
H$\alpha$ emitters is that they are {\it dusty red galaxies} 
(e.g. \citealt{wol05}; 2009).  If a star forming galaxy contains a 
significant amount of dust, it appears red due to the selective
extinction of bluer light, and in some cases, it becomes difficult
to be distinguished from passively evolving red galaxies. 
Furthermore, in the extremely dusty cases, even the H$\alpha$ lines
are heavily attenuated by dust and large corrections are required 
to obtain true star formation rates from the observed H$\alpha$ intensities
(e.g. \citealt{pog00}). 
This may explain the fact presented in the previous subsection 
that optically red MIPS-detected galaxies in the Abell\ 851 do not 
show significant H$\alpha$ emissions. 

An important point here is that, in contrast to the fact that
the optically red MIPS sources in the cluster core are not detected
in H$\alpha$, we do find many red H$\alpha$ emitters in the group environments.
If there are similar MIR-bright dusty red sources in groups, they should
{\it not} have been detected in H$\alpha$ in the groups, either.
Why do only the red star forming galaxies in groups have detectable
H$\alpha$ emissions? A possible answer would be:
(1) the star formation rates of these galaxies in groups are
significantly higher and have stronger intrinsic H$\alpha$ intensities, 
and/or (2) the mode of star formation in these systems is different 
from the MIPS sources in the cluster core (e.g. the location of star 
formation within galaxies and/or the geometry of dust extinction 
would be different).

\subsubsection{Strong starbursts or moderate star formation ?}

It has been reported that there is an excess of IR luminous star forming
galaxies in group-like environments in the distant Universe. 
For example, \cite{tra09} studied a supergroup environment at 
$z\sim 0.4$ in the MIR and showed an excess at the bright end of the IR 
luminosity function in group environment compared to cluster/field 
environments. Also, \cite{pog09a} studied the spectra
of galaxies in the EDisCS clusters at $z=$0.4--0.8 and showed that
the ``e(a)-type'' galaxies (those having a signature of dusty starburst
in their optical spectra) are most numerous in group environments.  
This excess in star forming activity of group galaxies may suggest that 
galaxy transition is actively taking place in group environments, 
and some galaxies with exceptionally high SFR may be 
detected in H$\alpha$. Also, recent MIR studies of 
distant clusters at $z\gsim 0.5$ found a large number of luminous
infrared galaxies (LIRGs) in the outskirts of clusters (e.g. 
\citealt{mar07}; \citealt{koy08}), and so it is likely
that such dusty starburst galaxies are included in our red H$\alpha$
emitter sample. In fact, \cite{koy10} discovered some red H$\alpha$
emitters in the outskirts of a $z=0.81$ cluster, and found that 
most of these red H$\alpha$ emitters are indeed MIR-detected LIRGs
with significant extinction at H$\alpha$ ($A_{\rm{H\alpha}} 
\gsim 3$ mag). \cite{gea09} conducted MIR spectroscopy for some 
LIRGs in the outskirt of the CL0024 cluster at $z=0.4$ reported 
in \cite{gea06}. They identified clear PAH 
emission features from the majority of their targets, suggesting 
that these LIRGs are really dusty starbursts (AGN contribution is small). 
Their MIR spectra resemble those of nuclear starbursts rather than normal
star formation in disks, and they 
proposed that such dusty starbursts seen in the outskirts of 
distant clusters are progenitors of the present-day cluster 
S0 galaxies (see also the discussion in Section~5.3.2). 

However, such extremely active galaxies tend to be more dusty, 
and it might be difficult to detect their H$\alpha$ emission. 
Also, the time-scale of dusty starbursts is not long, typically
$\lsim 1$Gyr. In our sample, we detect H$\alpha$ emission for 
$\sim$20--30\% of the red galaxies in the west clump and in the groups.
This fraction may be too large, if we assume that they are 
{\it all} short-lived dusty starbursts.  It is more likely that
a gentle mechanism which produces relatively long-lived red 
star forming galaxies is also at work in the group environment around
the Abell\ 851 cluster. \cite{wol09} studied dusty red galaxies 
in the complex structure of the Abell\ 901/902 clusters at $z=0.17$ 
with UV+MIR photometry, and found that their specific SFRs are 
systematically lower than blue star forming galaxies (see also 
\citealt{wol05} and \citealt{gal09} for their identification of 
dusty red galaxies).  
They concluded that the dusty red galaxies in Abell\ 901/902 are
``semi-passive'' rather than intense starbursts. They also proposed that
the origin of such red star forming galaxies is similar to that 
of ``passive spirals'' (galaxies with spiral morphology but without
on-going star formation activity; e.g. \citealt{pog99}, \citealt{got03}).
Although a direct comparison of their results with ours is difficult,
a fraction of the red H$\alpha$ emitters identified 
in our survey could be similar to the dusty red galaxies 
discussed in \cite{wol09} which are not strongly star-bursting galaxies
but moderately star-forming dusty galaxies.

\subsubsection{ Progenitors of the cluster S0 galaxies ?}

The existence of a large number of red star forming galaxies
and their location (i.e. preferably in groups) may indicate that
such populations are highly related to the transformation of
galaxies due to environmental effects, as there is a hint that
the group environment is a key place for galaxy truncation and 
the formation of S0 galaxies (see e.g. \citealt{wil08}; 2009). 
It has been reported that the fraction of S0 galaxies in cluster
cores dramatically increases from $z\sim 0.5$ to $z\sim 0$
(e.g. \citealt{dre97}; \citealt{des07}; \citealt{pog09b};
\citealt{jus10}), 
and it has been widely discussed in terms of the transformation 
from in-falling field spirals to cluster S0s as they assemble to 
clusters (e.g. \citealt{pog99}; \citealt{kod01a}). 
To reconstruct the S0 fraction in the nearby clusters, it may be 
required that galaxies are ``pre-processed'' in group environments
(e.g. Fujita 2004), and late-type spirals (S$_{\rm{cdm}}$) may be 
transformed into bulge-strong early spirals such as S$_{\rm{ab}}$ 
in the group environment (\citealt{kod01a}). Moreover, considering 
the high bulge-to-disk ratios of S0 galaxies, just a simple fading 
of spiral disks may not be sufficient to produce S0 galaxies
(\citealt{chr04}). The physical processes which can move the gas 
toward the galaxy center and can grow their bulges through a new 
star formation activity would be preferable.
The galaxy-galaxy interactions or harassment expected in groups
are the best candidates (see also the discussion in \citealt{mor07}).
In this context, many of the red H$\alpha$ emitters reported in this 
paper may be in the phase of growing bulges with significant star 
formation in the galactic central regions with moderate dust extinction.
However, the star formation activity is not as strong as a starburst,
and their H$\alpha$ emission is still visible through moderate
dust extinction.

\cite{mor07} studied passive spirals and young S0 galaxies in
two clusters at $z=0.4$ and $0.5$. They found that these transitional 
objects are preferentially seen in the infalling groups, and they 
also found that some of them are detected in UV. 
Such galaxies are probably similar 
populations to our red H$\alpha$ emitters. \cite{mor07} also examined the 
UV/optical colors of these objects and proposed that they are likely to be 
truncated with a long time-scale ($\gsim 1$Gyr), qualitatively 
consistent with our suggestion of long time-scales of the red 
H$\alpha$ emitters in group environment. 
On the other hand, \cite{mor07} found that the star
formation is more rapidly truncated in the cluster core environment 
(especially for more massive cluster with strong ICM). In our 
analysis, we did not find red H$\alpha$ emitters in the cluster
central region. This may be because the Abell\ 851 cluster is also 
a very rich cluster, and possibly red H$\alpha$ emitters cannot 
survive in the strong ICM in the cluster core (maybe their star 
formation is immediately shut off after entering the cluster core). 
In contrast, the MIPS-detected galaxies found in the cluster central
region would be more intense short-lived starbursts with stronger 
extinction, probably triggered by galaxy mergers as suggested by 
\cite{oem09}.

Unfortunately, we do not have MIR data (which are essential for
identifying obscured starbursts) and high-resolution HST data (which are also
essential to resolve galaxy morphology, mergers and localized star formation)
for the group environments where we find a large number of red 
H$\alpha$ emitters. These information should be powerful for
understanding the physical origin of the pre-processing working 
in the group environment, which may be closely related to the formation of 
cluster S0 galaxies. Combining these data with our wide-field 
H$\alpha$ imaging data will be clearly an important future work.

\subsubsection{Summary}

In summary, the red H$\alpha$ emitters are most commonly seen in galaxy 
groups around Abell\ 851. They are probably related to the physics 
of the ``pre-processing'' in group environments forming cluster S0 galaxies 
prior to entering the cluster core. The physical mechanisms mainly responsible 
for these populations are likely to be group-specific slow processes such as 
strangulation (e.g. \citealt{lar80}; \citealt{bek02}; \citealt{kaw08}) 
or harassment-like mechanisms (e.g. \citealt{moo99}; \citealt{mor07}). 
The latter might be more preferred if they are in the 
process of morphological change toward the more bulge-dominated 
early spirals at the same time. 
Strong dusty starbursts triggered by e.g. galaxy-galaxy
interactions or mergers may also be included in our red H$\alpha$ 
emitter sample, as we actually found such a population in the study 
of the $z\sim 0.8$ cluster (see \citealt{koy10}). 
However, it is currently difficult to quantify their relative 
contribution or its redshift evolution. The most demanding and 
promising next step is a wide-field observation of the infalling groups 
in the MIR/FIR with the future space IR missions. Dusty starbursts
should be bright in IR, while gradually fading star forming galaxies 
would not be so bright in IR. Such surveys will give us a critical 
information to identify the key process 
for the evolution of cluster galaxies.  

\section{Cluster Total Star Formation Rate}

Our narrow-band H$\alpha$ line survey is also capable to study
total star formation rate in clusters. The advantage of the narrow-band
survey is that we can trace star forming activity throughout the observed
field, and that we do not suffer from a sample selection bias or 
a completeness correction, which are inevitable for slit spectroscopy.
The wavelength range of the narrow-band filter used 
in this study is slightly offset from the actual redshift distribution of 
the cluster member galaxies in H$\alpha$ (see Fig.~1).
We correct for it in a statistical manner.
We use the velocity distribution of the spectroscopic members 
within 0.5$\times$R$_{200}$ from \cite{oem09}, and estimate the fraction 
of H$\alpha$ emission from the Abell\ 851 cluster that can be 
recovered by our NB921 filter.
The resultant correction factor to get the total star formation rate
turns out to be $\sim 1.54^{+0.08}_{-0.06}$ (the uncertainty is given as a
1$\sigma$ deviation derived from a bootstrap resampling of the spectroscopic
members).

\begin{figure}
 \begin{center}
 \rotatebox{0}{\includegraphics[width=9.0cm,height=9.0cm]{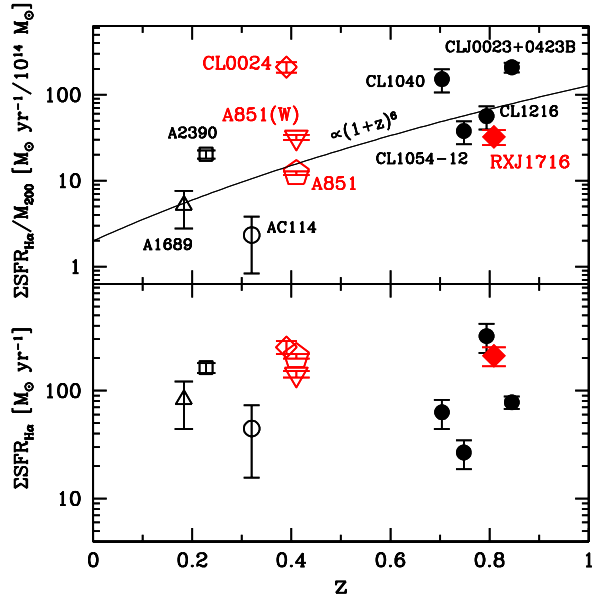}}
 \end{center}
 \vspace{-0.7cm}
\caption{ ({\it bottom}): The integrated total SFR(H$\alpha$) within
  0.5$\times$R$_{\rm{200}}$ as a function of redshift.
 Our measurements based on the Subaru data including our previous surveys
 are shown by red symbols. 
 ({\it top}): The integrated total SFR(H$\alpha$) normalized by the cluster
  mass M$_{200}$ as a function of redshift. 
\label{fig:total_SFR_vs_z}}
\end{figure}

Following the procedure in \cite{fin05} and \cite{koy10}, 
we sum up the star formation rates of H$\alpha$ emitters 
within 0.5$\times$R$_{200}$ from the cluster center 
to derive $\Sigma$SFR$_{\rm{H}\alpha}$, 
where R$_{200}$ is the radius within which the mean density is 200 
times larger than the critical density of the Universe (\citealt{car97}).
This allows us to directly compare our data with previous 
H$\alpha$-based cluster studies over a wide range in redshift, which
have been compiled by \cite{fin05}.
We also calculate the cluster mass, M$_{200}$, as in \cite{koy10}
and derive the cluster mass-normalized star formation rate (i.e.
$\Sigma$SFR$_{\rm{H}\alpha}$/M$_{200}$). The R$_{200}$ and M$_{200}$ 
are calculated in the following equations:
\begin{equation}
{\rm{R}}_{200} 
= 2.47 \times \frac{\sigma}{1000 \rm{km/s}} \frac{1}{\sqrt{\Omega
_{\Lambda} + \Omega _{\rm{M}} (1+z)^3}} \rm{Mpc}
\end{equation}
\begin{equation}
{\rm{M}}_{200} = 1.71 \times 10^{15} \left( \frac{\sigma}{1000 \rm{km/s}}  
\right)^3 \frac{1}{\sqrt{\Omega _{\Lambda} + \Omega _{\rm{M}} (1+z)^3}}
{\rm{M}}_{\odot},
\end{equation}
where $\sigma$ is the velocity dispersion of the cluster.
Using the velocity dispersion of the cluster core galaxies  
($\sigma = 1071$km/s; \citealt{oem09}) and the cosmological parameters
adopted in this paper, we derived R$_{200} =$2.13 Mpc and 
M$_{200} = 1.70\times 10^{15}M_{\odot}$. Then, using all 
galaxies with  $z'-$NB921$>$3$\sigma$, we derived 
$\Sigma$SFR$_{\rm{H}\alpha} = 208 \pm 10$M$_{\odot}$/yr 
and the mass-normalized star formation rate of 
$\Sigma$SFR$_{\rm{H}\alpha}/\rm{M}_{200} = 12.2 \pm 0.6 
M_{\odot}$yr$^{-1}$/$10^{14} M_{\odot}$.  
These values include the correction for the filter transmission 
curve as described above (i.e. a factor of $\sim 1.5$),
and the error-bars are given as a composite of the photometric 
errors and the uncertainty of the correction factor. 

Note that we include all galaxies 
with $z'-$NB921$>$3$\sigma$ that satisfy the same color selection 
criteria as in Fig.~\ref{fig:2color}. If we use only the secure
H$\alpha$ emitters as defined in this study
(i.e. $z'-$NB921$>$3$\sigma$ and $z'-$NB921$>$0.2), these values 
would become smaller by a factor of $\sim$2 (lower limit), and if we use 
all galaxies with $z'-$NB921$>$0, these values would increase by $\sim$25\%
(upper limit). Although this uncertainty is large, this does not affect 
our main conclusion. We expect that the uncertainty regarding the 
dust extinction may also be very large. In fact, we derive 
$\Sigma$SFR(IR)$\sim 370$M$_{\odot}$/yr by just summing up the 
IR-derived SFRs of the 16 MIPS-detected spec-$z$ members listed in 
Table 2, which is clearly larger than $\Sigma$SFR(H$\alpha$) derived 
above. Although we use a constant 1 mag extinction correction in this 
paper to make a fair comparison with the values in other works in the 
literature, it is also essential to study such obscured/hidden 
activities more in detail with the MIR--FIR observations (see some 
examples of such IR studies of the global evolution of star forming 
activities in clusters by e.g. Geach et al.\ 2006; Bai et al.\ 2007; 
Koyama et al.\ 2010; Haines et al.\ 2009a; Chung et al.\ 2011). 

We plot in Fig.~\ref{fig:total_SFR_vs_z} the values derived above 
and compare them with those for other clusters in the literature 
all based on H$\alpha$ (see \citealt{koy10} and references therein). 
This plot clearly shows that the mass-normalized star formation rate 
of Abell\ 851 is located on the general evolutionary trend with 
redshift, following approximately $\propto (1+z)^6$. 
The lower limit in SFR that is used to calculate $\Sigma$SFR
is slightly different from cluster to cluster, but it is within the 
range of 0.1--1 M$_{\odot}$/yr (dust-free). Since the integrated 
SFRs are dominated by galaxies with strong star formation,   
we do not correct for this effect (see also \citealt{fin04}; 
\citealt{kod04}). The strong evolution of star forming activity
in clusters presented above is well consistent with our previous study 
in \cite{koy10}, and this trend is much more significant than the 
trend of a decrease in the specific star formation rates of 
{\it individual} galaxies, $\propto (1+z)^3$ shown by e.g. 
\cite{yos06}; \cite{zhe07} (see also the H$\alpha$-based
studies of cosmic star formation history by e.g. \citealt{sob09}; 
\citealt{wes10}; \citealt{dal10}). 
Therefore, our result also suggests that cluster environment 
does indeed accelerate the quenching of activities in galaxies.  

\begin{figure}
 \begin{center}
 \rotatebox{0}{\includegraphics[width=9.0cm,height=9.0cm]{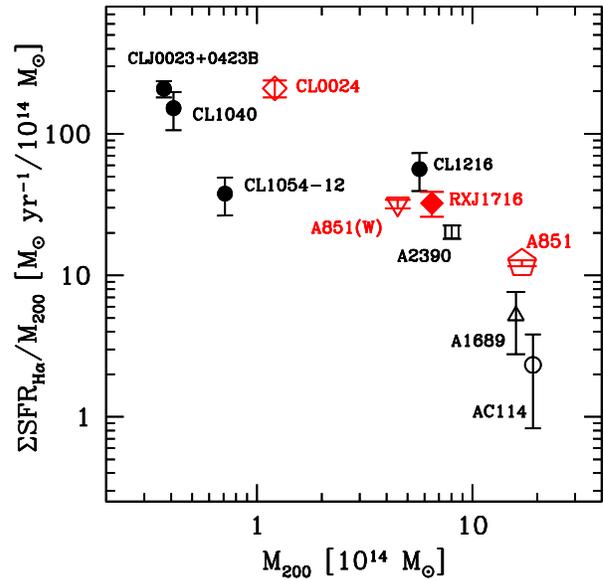}}
 \end{center}
 \vspace{-0.6cm}
\caption{ The integrated total SFR(H$\alpha$) normalized by the cluster
  mass as a function of cluster mass, M$_{200}$. The symbols are
 the same as Fig.~11. Open and filled symbols indicate the low-$z$
 samples ($z\lsim 0.4$) and the high-$z$ samples ($z\gsim 0.6$), 
 respectively.  
\label{fig:total_SFR_vs_mass}}
\end{figure}

We note that this redshift dependence of the cluster mass-normalized 
star formation rates (or its scatter) may be related to the cluster mass 
growth (e.g. \citealt{fin05}; \citealt{koy10}). As presented in \cite{koy10},
more massive clusters exhibit lower 
$\Sigma$SFR$_{\rm{H}\alpha}/\rm{M}_{200}$
(see also \citealt{hom05}; \citealt{bai09}). This may also be related 
to the fact that the fraction of star forming galaxies is a decreasing 
function of cluster mass (e.g. \citealt{pog06}, but see a 
different result by \citealt{hai09b}). 
In fact, a large scatter between clusters is clearly visible 
even at a fixed redshift in the top panel of Fig.~\ref{fig:total_SFR_vs_z}. 
In particular, the value of Abell\ 851 presented in this paper 
is significantly different from that of CL0024 cluster at a similar
redshift (see Fig.~\ref{fig:total_SFR_vs_z}). 
However, Abell\ 851 is $\sim$1 order of magnitude more massive 
than CL0024, and both clusters still lie along the general trend 
of decreasing $\Sigma$SFR$_{\rm{H}\alpha}/\rm{M}_{200}$ 
with increasing cluster mass (Fig.~\ref{fig:total_SFR_vs_mass}). 
Furthermore, we also calculated these values for the West Clump 
of Abell\ 851 in the same way as for the main cluster, adopting 
the R$_{200}$ and M$_{200}$ taken from the preliminary result 
of the optical spectroscopy by \cite{nak11}. We plotted these 
values for the west clump in Figs.~11 and 12. Interestingly, 
the west clump of Abell\ 851 is more massive than CL0024 cluster 
and has a lower value of $\Sigma$SFR$_{\rm{H}\alpha}/\rm{M}_{200}$, 
indicating the importance of the mass of the structure in determining 
the star forming activity in the system.

We should comment that this simultaneous trend of $\Sigma$SFR vs. 
cluster mass and $\Sigma$SFR vs. redshift makes it difficult to derive 
any evolutionary trend in star formation activities in clusters.
Our fitting of a form $\Sigma$SFR$\propto (1+z)^m \times {\rm{M_{cl}}}^n$ 
to the data shown in Figs 11 and 12 yielded $m=1.2 \pm 2.0$ and 
$n = -0.75 \pm 0.2$. However, these numbers should be taken as very 
preliminary because they are based on a small sample. It should be 
also noted that, as shown in Fig.12, the low-$z$ clusters (open symbols) 
studied so far tend to be more massive, while the high-$z$ clusters 
(filled symbols) tend to be less massive. This may have led to an 
apparently stronger evolutionary trend than the actual one. It is 
thus critically important to collect a large number of distant 
clusters covering various masses and redshifts with future large 
cluster surveys.

\section{Summary and Conclusions}

We performed a panoramic narrow-band H$\alpha$ emitter survey for an 
intermediate redshift cluster Abell\ 851 at $z=0.41$ with Suprime-Cam
on the Subaru Telescope. This is one of the most distant wide-field 
H$\alpha$ emitter surveys ever made for clusters. After selecting the 
H$\alpha$ emitters with color excess in $z'-$NB921 and the broad-band 
color information, we first map out the spatial distribution of the H$\alpha$ 
emitters around Abell\ 851. The H$\alpha$ emitters are found throughout 
the observed 27$'\times$27$'$ field, suggesting that the large-scale structures
identified by our previous optical imaging survey are really located
at the cluster redshift and physically associated with the main cluster. 
The fraction of H$\alpha$ emitters is a strong function of environment
and shows a clear decline toward the cluster central region.

The most important result of this study is the color variation 
of H$\alpha$ emitters and its clear environmental dependence. 
The majority of the H$\alpha$ emitters are blue galaxies with $B-I<2$, 
but we find that $\sim$15\% of the H$\alpha$ emitters have red colors.
Such red H$\alpha$ emitters are preferentially found in the group 
environment relatively far away from the cluster core.
Our survey shows that $\sim$20--30\% of H$\alpha$ emitters in 
the west clump and the groups have red colors with $B-I>2$, while such
red emitters are very rare in and immediately outside of the cluster 
core ($\lsim$5\%). 
Some of the red H$\alpha$ emitters might be dusty starbursts with 
significant star formation produced via e.g. galaxy-galaxy interactions 
or mergers.  However, it would be unlikely that {\it all} the red
H$\alpha$ emitters are dusty starbursts, because most of the optically 
red MIR-detected galaxies found in the cluster central region are 
{\it not} detected in H$\alpha$ (and also because the number of red
H$\alpha$ emitters is apparently too large for their expected 
short life time). Therefore, it would be reasonable to interpret 
that the red H$\alpha$ emitters in group environment are generated 
by relatively gentle and slow processes which are different from 
those producing MIR sources in the core of 
Abell\ 851. Also, a large number of the red H$\alpha$ emitters found in the 
groups might be closely related to the ``pre-processing'' contributing to 
the formation of bulge-dominated cluster S0 galaxies before entering 
the cluster core region.
 
We also derived the cluster total star formation rate and the cluster
mass-normalized star formation rate by summing up the star formation
rates of individual galaxies within 0.5$\times$R$_{200}$
of the main cluster and the west clump.
We confirmed a general trend that the mass-normalized star 
formation rates increase sharply toward distant clusters following
approximately $\propto (1+z)^6$, which turned out to be remarkably consistent 
with our previous studies, although there is a large scatter
between clusters. This scatter may be related to the variation 
of the cluster mass at each epoch. More massive clusters have 
lower star formation activity inside them, suggesting that the
cluster mass growth is related to the suppression of the
star formation activity in galaxies. 

\vspace{5mm}

\acknowledgments

We thank the anonymous referee for relevant comments and suggestions. 
Y.K. also thank Dr. Kentaro Motohara, Prof. Kotaro Kohno, Prof. Masanori
Iye, Prof. Yuzuru Yoshii, and Prof. Nobuo Arimoto for their careful
reading of the manuscript and helful comments. 
The broad-band and narrow-band imaging data used in this paper are 
collected at the Subaru Telescope, which is operated by the National 
Astronomical Observatory of Japan (NAOJ). Y.K. acknowledge the support 
from the Japan Society for the Promotion of Science (JSPS) through 
JSPS research fellowships for Young Scientists. 
This work was financially supported in part by a Grant-in-Aid for the
Scientific Research (Nos.\, 18684004; 21340045) by the Japanese 
Ministry of Education, Culture, Sports and Science. 

{\it Facilities:} \facility{Subaru} .

\end{document}